\documentclass{svproc}

\usepackage{url}

\usepackage{epsfig}
\usepackage{subfigure}
\usepackage{calc}
\usepackage{amssymb}
\usepackage{amstext}
\usepackage{amsmath}
\usepackage{multicol}
\usepackage{xcolor}
\usepackage{textcomp}

\usepackage{cite}
\usepackage{graphicx}
\graphicspath{ {./figures/} }

\usepackage{mathrsfs}
\usepackage{verbatim}
\usepackage{url}

\DeclareMathOperator*{\mathsup}{sup}
\DeclareMathOperator\erf{erf}

\newcommand{\RR}{\ensuremath{\mathbb{R}}}

\begin{document}
\mainmatter
\title{Quantifying Robotic Swarm Coverage} %  \subtitle{Preparation of Camera-Ready Contributions to SCITEPRESS Proceedings} }
\titlerunning{Quantifying Robotic Swarm Coverage}

\author{Brendon Anderson\inst{1} \and Eva Loeser\inst{2} \and Marissa Gee\inst{3} \and Fei Ren\inst{4} \and Swagata Biswas\inst{5} \and Olga Turanova\inst{6} \and Matt Haberland\inst{7} \and Andrea L. Bertozzi\inst{8}}

\authorrunning{Anderson et al.}
\institute{UC Berkeley, Department of Mechanical Engineering, Berkeley, CA 94720
\and
UC San Diego,  Department of Mathematics, San Diego, CA 92093
\and
Cornell University, Center for Applied Mathematics, Ithaca, NY 14853
\and
UC Berkeley, Industrial Engineering and Operations Research, Berkeley, CA 94720
\and
University of Edinburgh, School of Mathematics, Edinburgh,
Scotland, EH9 3FD
\and
Institute for Advanced Study, School of Mathematics, Princeton, NJ 08540
\and
Cal Poly, BioResource and Agricultural Engineering, San Luis Obispo, CA 93407\\
\email{mhaberla@calpoly.edu}
\and
UCLA, Department of Mathematics, Los Angeles, CA 90095}

\maketitle 

\abstract{In the field of swarm robotics, the design and implementation of spatial density control laws has received much attention, with less emphasis being placed on performance evaluation. This work fills that gap by introducing an error metric that provides a quantitative measure of coverage for use with any control scheme. The proposed error metric is continuously sensitive to changes in the swarm distribution, unlike commonly used discretization methods. We analyze the theoretical and computational properties of the error metric and propose two benchmarks to which error metric values can be compared. The first uses the realizable extrema of the error metric to compute the relative error of an observed swarm distribution. We also show that the error metric extrema can be used to help choose the swarm size and effective radius of each robot required to achieve a desired level of coverage. The second benchmark compares the observed distribution of error metric values to the probability density function of the error metric when robot positions are randomly sampled from the target distribution. We demonstrate the utility of this benchmark in assessing the performance of stochastic control algorithms. We prove that the error metric obeys a central limit theorem, develop a streamlined method for performing computations, and place the standard statistical tests used here on a firm theoretical footing. We provide rigorous theoretical development, computational methodologies, numerical examples, and MATLAB code for both benchmarks.}
% \abstract{This paper studies a generally applicable, sensitive, and intuitive error metric for performance assessment of robotic swarm controllers. Inspired by vortex blob numerical methods, this error metric overcomes the shortcomings of a common strategy based on spatial discretization, and unifies other continuous notions of coverage. We present two benchmarks against which to compare the error metric value of a given swarm configuration: non-trivial bounds on the error metric, and the  probability density function of the error metric when robot positions are sampled at random from the target swarm distribution. We give rigorous results showing that this probability density function of the error metric obeys a central limit theorem, allowing for more efficient numerical approximation and standard statistical tests. For both of the benchmarks developed here, we present supporting theory, computation methodology, examples, and MATLAB implementation code.

\section{\uppercase{Introduction}}
\label{sec:introduction}

\noindent Much of the research in swarm robotics has focused on determining control laws that elicit a desired group behavior from a swarm \cite{brambilla2013swarm}, and less attention has been placed on methods for evaluating the performance of these controllers quantitatively. Both \cite{brambilla2013swarm} and \cite{cao1997cooperative}  point out the lack of developed performance metrics for assessing and comparing swarm behavior, and \cite{brambilla2013swarm} notes that existing performance metrics are often too specific to the task being studied to be useful in comparing performance across controllers. 

This extended version of the authors' paper \cite{ICINCO2018} studies an error metric that evaluates one common desired swarm behavior: distributing the swarm according to a prescribed spatial density. Here, we delve deeper into the analysis of the properties of the error metric. Further, we include new examples, including a continuous target distribution to complement the original piecewise constant distribution, in order to better illustrate the utility of our methods.

In many applications of swarm robotics, the swarm must spread across a domain according to a target distribution in order to achieve its goal. Some examples are in surveillance and area coverage \cite{bruemmer2002robotic,hamann2006analytical,howard2002mobile,schwager2006distributed}, achieving a heterogeneous target distribution \cite{elamvazhuthi2016coverage,berman2011design, demir2015decentralized,shen2004hormone,elamvazhuthi2015optimal, CMK, Zhao}, and aggregation and pattern formation \cite{soysal2006macroscopic,spears2004distributed, reif1999social,sugihara1996distributed, Pimenta2008, Pimenta2013}. Despite the importance of assessing performance, some studies such as \cite{shen2004hormone, sugihara1996distributed, Pimenta2008, Pimenta2013, Zhao} rely only on qualitative methods such as visual comparison. Others present performance metrics that are too specific to be used outside of the specific application, such as measuring cluster size in \cite{soysal2006macroscopic}, distance to a pre-computed target location in \cite{schwager2006distributed, CMK}, and area coverage by tracking the path of each agent in \cite{bruemmer2002robotic}. Ergodicity has also been used as a notion of coverage error \cite{MM, ASC}, but it quantifies only the time-average statistics of the robot trajectories rather than the effective coverage of the swarm at an instant in time.
 In \cite{reif1999social}  an $L^2$ norm of the difference between the target and achieved swarm densities is considered, but the notion of achieved swarm density is particular to the controllers under study. 
These existing works do not provide methodology for carrying out computations with the notions they introduce. Moreover, they do not give guidance on how to interpret the values produced by the error metrics they define. Our work seeks to fill these gaps.

We develop and analyze an error metric that quantifies how well a swarm achieves a prescribed spatial distribution. Our method is independent of the controller used to generate the swarm distribution, and thus has the potential to be used in a diverse range of robotics applications.  
In \cite{li2017decentralized, zhang2017performance, EA}, error metrics similar to the one presented here are used, but their properties are not discussed in sufficient detail for them to be widely adopted. 
In particular, although the error metric that we study always takes values somewhere between 0 and 2, these values are, in general, not achievable for an arbitrary desired distribution and a fixed number of robots. %How then, in general, is one to judge whether the value of the error metric, and thus the robot distribution, achieved by a given swarm control law  ``good'' or not? 
Therefore, one needs a better understanding of the finer properties of the error metric in order to judge whether its values (and hence the performance of the underlying controller) are ``good" or not.
We address this by studying two benchmarks, 
\begin{enumerate}
\item the \emph{extrema} of the error metric, and
\item the \emph{probability density function} (PDF) of the error metric  when robot positions are sampled from the target distribution,
\end{enumerate}
which were first proposed in \cite{li2017decentralized}. Using tools from nonlinear programming for (1) and rigorous probability results for (2), we put each of these benchmarks on a firm foundation.   In addition, we provide MATLAB code for performing all calculations at \url{https://git.io/v5ytf}. Thus, by using the methods developed here, one can assess the performance of a given controller for any target distribution by comparing the error metric value of robot configurations produced by the controller against benchmarks (1) and (2).

Our paper is organized as follows. Our main definition, its basic properties, and a comparison to common discretization methods is presented in Section \ref{sec:error}. Then,  Section \ref{sec:extrema} and Section \ref{sec:errorpdf} are devoted to studying (1) and (2), respectively.  
We suggest future work in Section \ref{sec:future_work} and conclude in Section \ref{sec:conclusion}.

\section{\uppercase{Quantifying Coverage}}
\label{sec:error}

\noindent One difficulty in quantifying swarm coverage is that the target density within the domain is often prescribed as a continuous (or piecewise continuous) function, yet the swarm is actually composed of a finite number of robots at discrete locations. %A common approach for comparing the actual positions of robots to the target density function begins by discretizing the domain (e.g. \cite{berman2011design, demir2015decentralized}). We demonstrate the pitfalls of this in Subsection \ref{sec:naive}. 
%Another possible route (the one we take here) 
The approach we take here is to use the robot positions to construct a continuous function that represents coverage (e.g.  \cite{HS,ZC,ASC, EA, Pimenta2013}). It is also possible to use a combination of the two methods, as in \cite{Cortes}. 

The method we present and analyze is inspired by vortex blob numerical methods for the Euler equation and the aggregation equation (see \cite{CB} and the references therein). There are also similarities between our approach and the numerical methods known as smoothed particle hydrodynamics (SPH), which have also been applied in robotics  \cite{Pimenta2008, Pimenta2013, Zhao}. One main idea behind vortex methods and SPH is to use particles to approximate a continuous function that represents the fluid. We apply this idea but with the opposite aim: namely, we represent discrete points (the robots' positions) with a continuous function. A similar strategy was alluded to in \cite{EA} and  used in \cite{li2017decentralized, zhang2017performance} to measure the effectiveness of a certain robotic control law, but to our knowledge, our work here and in \cite{ICINCO2018} is the first to develop any such method in a form sufficiently general for common use.  % . Our definition of error metric below is the time-independent version of that in \cite{li2017decentralized}. To our knowledge, this paper the first generalization of this error metric

This section is devoted  to our definition of the error metric and to its basic properties and computational considerations. We also provide a contrast between our method and another common way to measure error, namely, by discretizing the domain  (e.g. \cite{berman2011design, demir2015decentralized}) in Subsection \ref{sec:naive}. Finally, in Subsection \ref{sec:two ex}, we present the setup for the two main examples that we will use throughout the paper.

\subsection{Preliminaries}

%\textcolor{red}{This subsection is dedicated to the definition of the error metric that we will use throughout the paper and to its basic properties and related concepts.} 
We are given a bounded region\footnote{We present our definitions for any number of dimensions $d\geq 1$ to demonstrate their generality. However, in  the latter sections of the paper, we restrict ourselves to $d=2$, a common setting in ground-based applications.}  
$\Omega \subset \RR^d$, a desired robot distribution $\rho:\Omega \rightarrow (0, \infty)$ satisfying $\int_\Omega \rho(z)\, dz=1$, and $N$ robot positions $x_1, \dots, x_N\in \Omega$.  
To compare the discrete points $x_1, \dots, x_N$ to the function $\rho$, we place a ``blob'' of some shape and size at each point $x_i$. The blob or \emph{robot blob function} can be any function $K:\RR^d \rightarrow \RR$ that  is non-negative on $\Omega$ and satisfies $\int_{\RR^d}K(z)dz = 1$. In other contexts the blob function is referred to as a ``kernel'' \cite{WandJones, Pimenta2008, Zhao}. Although more general blob shapes may be useful to represent properties of a robot's sensing and/or manipulation capabilities, it is often natural to use a $K$ that is radially symmetric, decreases along radial directions, and enjoys certain integrability properties. We discuss these issues further in Subsection \ref{subsec:K}.

One choice of $K$ could be a scaled indicator function, for instance, a function of constant value within a disc of radius $1$ and $0$ elsewhere. This is an appropriate choice when a robot is considered to either perform its task at a point or not, and where there is no notion of the degree of its effectiveness. For the remainder of this paper, however, we usually take $K$ to be the Gaussian 
\[
G(z)=\frac{1}{2\pi}\exp\left(-\frac{|z|^2}{2}\right),
\]
which is useful when the robot is most effective at its task locally and to a lesser degree some distance away. %\textcolor{red}{A list of other common choices of $K$ can be found in \cite[Chapter 2]{WandJones}.}
A list of other common choices of $K$ and different ways of constructing multivariate blobs from single variable functions can be found in \cite[Chapter 2, Chapter 4]{WandJones}.

To formulate our definition, we need one more parameter, a positive number $\delta$, which we call \emph{blob radius}. 
 % that controls how far the effective area (or inherent positional uncertainty) of the robot extends. 
 We then define $K^\delta$ as,
\begin{equation}
\label{def:Kdelta}
K^\delta(z)=\frac{1}{\delta^d} K\left(\frac{z}{\delta}\right).
\end{equation}
We point out that this rescaling preserves the $L^1$ norm of $K$, so that $\int_{\RR^d} K^\delta(z)\, dz=1$ holds for all $\delta>0$.

The shape of $K$ and the blob radius  $\delta$ have two  physical interpretations as:
\begin{itemize}
\item the area in which an individual robot performs its task, or
\item inherent uncertainty in the robot's position.
\end{itemize}
Either of these interpretations (or a combination of the two) can be invoked to make a meaningful choice of these parameters.

 To define the \emph{swarm blob function} $\rho_N^\delta$, we place a blob $G^\delta$ at each robot position $x_i$, sum over $i$ and renormalize,  yielding,
\begin{equation}
\label{eqn:normalized_blob_function}
\rho_N^\delta(z;x_1,..., x_N) = \frac{\sum_{i=1}^N G^\delta(z-x_i)}{\sum_{i=1}^N \int_\Omega G^\delta(z-x_i)\, dz}.
\end{equation}
For brevity, we usually write $\rho_N^\delta(z)$ to mean $\rho_N^\delta(z;x_1,..., x_N)$. 
This swarm blob function gives a continuous representation of how the discrete robots are distributed. 
Note that each integral in the denominator of (\ref{eqn:normalized_blob_function}) approaches $1$ if $\delta$ is small or all robots are far from the boundary, so that we have,
\begin{equation}
\label{eqn:rho with N}
\rho_N^\delta(z) \approx \frac{1}{N}\sum_{i=1}^N G^\delta(z - x_i).
\end{equation}
 Moreover, $\rho_N^\delta(z; x_1,..., x_N)$ is a smoothed version of $\displaystyle \sum_{i=1}^N \delta_{x_i}(z)$, which represents the physical robot positions (here, $\delta_{x_i}(z)$ denotes the Dirac mass at $x_i$). Indeed, 
\begin{equation}
\label{eq:delta masses}
\lim_{\delta\rightarrow 0} \rho_N^\delta(z; x_1,..., x_N) = \frac{1}{N}\sum_{i=1}^N \delta_{x_i}(z)
\end{equation}
in the sense of distributions \cite{brezis2010functional}. This point of view gives yet another interpretation of $\rho_N^\delta$.

We now introduce our notion of error:
\begin{definition}
\label{def:error}
The \emph{error metric} $e_N^\delta $ is defined as:
\begin{equation}
\label{eqn:instantaneous_error_metric}
e_N^\delta(x_1,...,x_N)=\int_\Omega \left| \rho_N^\delta(z) - \rho(z) \right| dz .
\end{equation}
\end{definition}
This definition, including the definitions of $K^\delta$ and $\rho_N^\delta$, was introduced in this context by the authors in \cite{ICINCO2018}. The relation (\ref{eqn:rho with N}) appeared there as well.

%We now define our notion of error, which we refer to as the \emph{error metric}: 
%\begin{equation}
%\label{eqn:instantaneous_error_metric}
%e_N^\delta(x_1,...,x_N)=\int_\Omega \left| \rho_N^\delta(z) - \rho(z) \right| dz .
%\end{equation}
%We often write this as $e_N^\delta $ for brevity. 

\subsubsection{Remarks and Basic Properties}
\label{sec:error_properties}
Our error is  defined as the $L^1$ norm of the difference between the swarm blob function $\rho^\delta_N$ and the desired robot distribution $\rho$. One could use another $L^p$ norm; however, $p=1$ is a standard choice in applications that involve particle transportation and coverage such as \cite{zhang2017performance, EA}. Moreover, the $L^1$ norm has a key property:  for any two integrable functions $f$ and $g$, 
\[
\int_\Omega\left|f-g\right| \, dz = 2\mathsup_{B\subset \Omega} \left|\int_B f \, dz - \int_B g \,  dz\right|.
\]
The other $L^p$ norms do not enjoy this property \cite[Chapter 1]{DG}. Consequently, by measuring $L^1$ norm on $\Omega$, we are also bounding the error we make on any particular subset, and, moreover, knowing the error on ``many'' subsets gives an  estimate of the total error. This means that by using the $L^1$ norm we capture the idea  that discretizing the domain provides a measure of error, but avoid the pitfalls of discretization methods described in Subsection \ref{sec:naive}.

Studies in optimal control of swarms often use the $L^2$ norm due to the favorable inner product structure \cite{zhang2017performance}.  We point out that the $L^1$ norm is bounded from above by the $L^2$ norm: indeed, according to the Cauchy-Schwarz inequality, for any function $f$ we have, 
\[
\int_\Omega f \, dz \leq |\Omega| \left(\int_\Omega f^2\, dz\right)^{1/2} ,
\]
where  $|\Omega|$ denotes the area of the bounded region $\Omega$.  Thus, if an optimal control strategy controls the $L^2$ norm, then it will also control the error metric we present here.

Last, we note that for any $\Omega$, $\rho$, $\delta$, $N$, and $(x_1,...,x_N)$, we have $0\leq e_N^\delta \leq 2$. This was established in Proposition 2.1 of \cite{ICINCO2018}. 
The theoretical minimum of $e^\delta_N$  can only be approached for a general target distribution when $\delta$ is small and $N$ is large, or in the trivial special case when the target distribution is exactly the sum of $N$ Gaussians of the given $\delta$, motivating the need to develop benchmarks (1) and (2).

\subsubsection{Variants of the Error Metric}
The notion of error in Definition \ref{def:error} is suitable for tasks that require good instantaneous coverage. For tasks that involve tracking average coverage over some period of time (and in which the robot positions are functions of time $t$), an alternative ``cumulative'' version of the error metric is
\begin{equation}
\label{eqn:cumulative_error_metric}
\int_\Omega \left| \frac{1}{M}\sum_{j = 1}^M{\rho_N^\delta(z,t_j)} - \rho(z) \right| dz
\end{equation}
for time points $j = 1, \dots, M$. This was defined in the authors' \cite{ICINCO2018}, and is similar to the metric used in \cite{zhang2017performance}. It may also be likened to the approach of \cite{ASC}, which expresses the average over time using an integral rather than finite sum and measures the difference from the target distribution using the Kullback-Leibler divergence and Bhattacharyya distance instead of an $L^1$ norm. Although the cumulative error metric (\ref{eqn:cumulative_error_metric}) is, in general, distinct from the instantaneous version of (\ref{eqn:instantaneous_error_metric}), note that the extrema and  PDF of this cumulative version can be calculated  in the same way as the extrema and PDF of the instantaneous error metric  with $MN$ robots. Therefore, in subsequent sections we restrict our attention to the extrema and PDF of the instantaneous formulation without loss of generality.

In addition, \cite{zhang2017performance} considers a one-sided notion of error, in which a scarcity of robots is penalized but an excess is not, that is,
\[
\hat{e}_N^\delta = \int_{\Omega^-} \left| \rho^\delta_N(z) - \rho(z) \right| \, dz, 
\]
where $\Omega^-:=\{z|\rho_N^\delta(z)\leq \rho(z)\}$.  The definition of $\hat{e}_N^\delta$, which appears in \cite{ICINCO2018}, is particularly useful in conjunction with the choice of $K^\delta$ as a scaled indicator function, as $\hat{e}_N^\delta$ becomes a direct measure of the deficiency in coverage of a robotic swarm. For instance, given a swarm of surveillance robots, each with observational radius $\delta$, $\hat{e}_N^\delta$ is the percentage of the domain not observed by the swarm.\footnote{The notion of ``coverage'' in \cite{bruemmer2002robotic} might be interpreted as $\hat{e}_N^\delta$ with $\delta$ as the width of the robot. There, only the time to complete coverage ($t$ such that $\hat{e}_N^\delta(z;x_1(t), ..., x_n(t)) = 0$) was considered.} 

Remarkably, $\hat{e}_N^\delta$ and $e_N^\delta$ are related by $e_N^\delta = 2\hat{e}_N^\delta$. This was established by the authors in Proposition 2.2 of \cite{ICINCO2018}. This relationship implies that $e_N^\delta$ enjoys the interpretation of being a measure of deficiency, and also allows the techniques introduced here to be directly applied to $\hat{e}_N^\delta$.

\subsection{Calculating $e_N^\delta$}
\begin{figure}[ht]
\begin{center}
\includegraphics[width=.75 \linewidth]{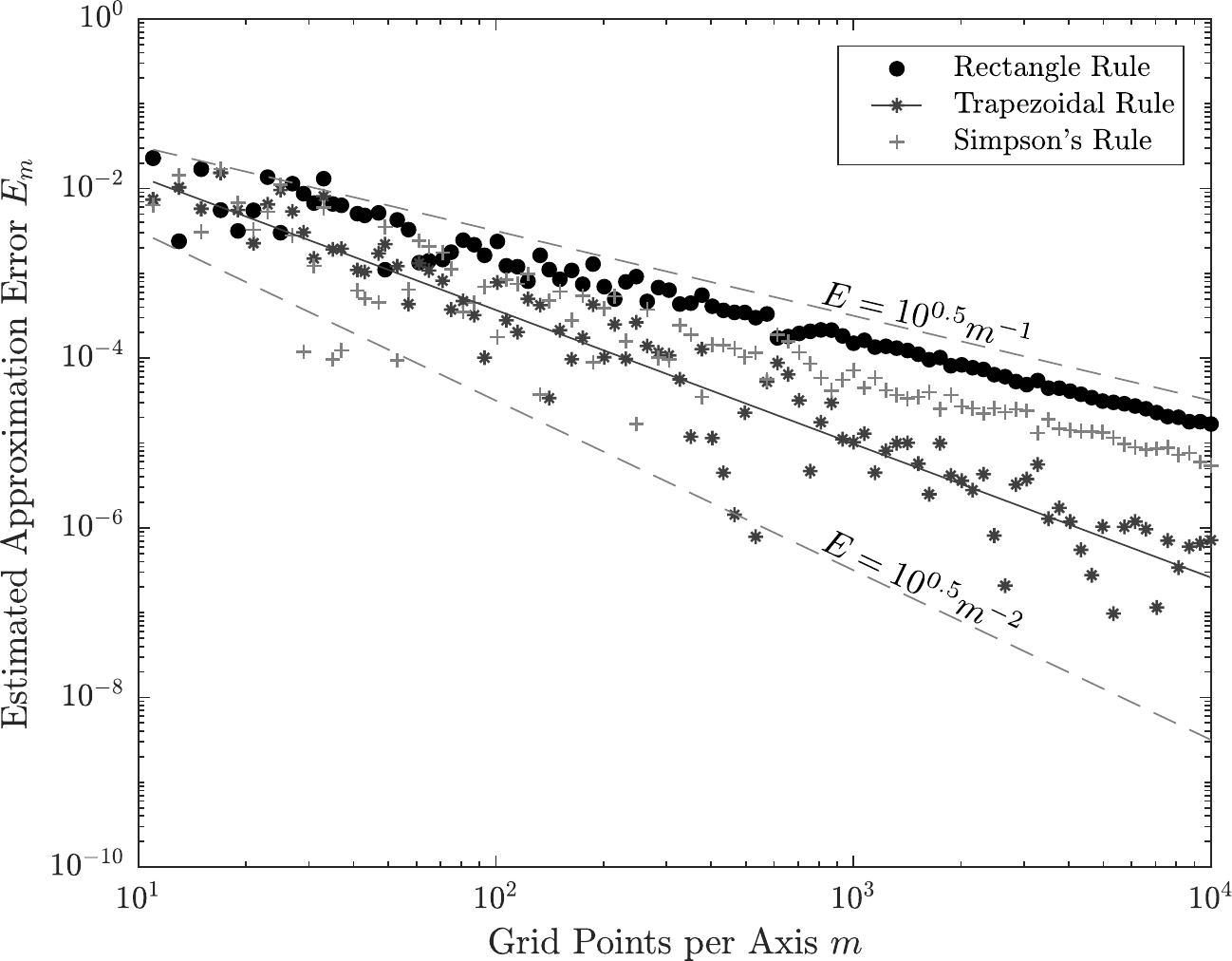}
\caption{Estimated error $E_m = \left| e_N^\delta - \tilde{e}_N^\delta  \right|$ as the number of grid points per axis $m$ increases. As the exact value of $e_N^\delta$ is unknown, we use the estimate provided by the MATLAB function \texttt{integral2} using the \texttt{\textquotesingle iterated\textquotesingle} method and a relative tolerance of $10^{-8}$. The solid line is a best fit function of the form $10^a m^b$ for the trapezoidal rule, where $a = 0.2782$ and $b = 1.577$ were found using the MATLAB Curve Fit app. The particular configuration of robots is the optimal arrangement found (see Section \ref{subsec:swarm design}) for the ring distribution from Definition \ref{defn:ring} with $N=200$ robots and $\delta = 2$in.}
\label{fig:emi}
\end{center}
\end{figure}

In practice, the integral in (\ref{eqn:instantaneous_error_metric}) can rarely be carried out analytically, primarily because the integral needs to be separated into regions for which the quantity $\rho_N^\delta(z) - \rho(z)$ is positive and regions for which it is negative, the boundaries between which are usually difficult to express in closed form. Hence we must rely on numerical integration, or quadrature. Here we study the magnitude $E_m$ of the difference between the true value of the error metric $e_N^\delta$ and an approximation $\tilde{e}_N^\delta$ as the number of grid points per axis $m$ increases for three elementary quadrature rules: the rectangle rule, the trapezoidal rule, and Simpson's rule.  As quadrature convergence rate proofs typically depend on smoothness of the integrand \cite{cruz2002sharp}, we did not expect all of these rules to achieve their nominal rates of convergence. Indeed, in Figure \ref{fig:emi} we see that the trapezoidal rule converges as $O(m^{-1.58})$ instead of $O(m^{-2})$ and Simpson's rule converges as $O(m^{-1})$ instead of $O(m^{-4})$. As our applications do not require extremely accurate estimates of the error metric, other error metric values reported throughout the paper have been calculated using the rectangle rule with a moderate number of grid points. For applications that demand increased accuracy, we recommend an adaptive scheme such as that employed by MATLAB's \texttt{integral2} function.

\subsection{The Pitfalls of Discretization}
\label{sec:naive}
Before concluding this section, we analyze a measure of error that involves discretizing the domain  and  show  that the values produced by this method are strongly dependent on a choice of discretization.  % that is not readily made based on physical arguments. 
In particular, this error approaches its theoretical minimum when the discretization is too coarse and its theoretical maximum when the discretization is too fine, regardless of robot positions.

Discretizing the domain means dividing  $\Omega$  into $M$ disjoint regions $\Omega_i \subset \Omega$ such that $\bigcup_{i=1}^{M}\Omega_i = \Omega$. 
Within each region, the desired proportion of robots is the integral of the target density function within the region $\int_{\Omega_i} \rho(z) dz$. Using $N_i$ to denote the observed number of robots in $\Omega_i$, we can define an  error metric as
\begin{equation}
\label{eqn:naive_error}
\mu = \sum_{i=1}^M \left| \int_{\Omega_i} \rho(z)\, dz - \frac{N_i}{N} \right|.
\end{equation}
This exact definition appeared in the authors' \cite{ICINCO2018}, and is based on discrete error metrics used in practice, for instance in \cite{berman2011design, demir2015decentralized}. It is easy to check that $0\leq \mu\leq 2$ always holds. 
One advantage of this approach is that $\mu$ is very easy to compute, but there are two major drawbacks. 

%\subsubsection{Choice of Domain Discretization}
First, the choice for domain discretization is not unique, and this choice can dramatically affect the  value of $\mu$.  Indeed, if $M=1$ then $\mu=0$. This follows directly from the definition of $\mu$ and appears in \cite{ICINCO2018} as Proposition 2.3. On the other hand, as the discretization becomes finer, the error approaches its maximum value 2, regardless of the robot positions. More precisely:
\begin{proposition} Suppose  the robot positions are distinct\footnote{This is reasonable in practice as two physical robots cannot occupy the same point in space. In addition, the proof can be modified to produce the same result even if the robot positions coincide.} and  the regions $\Omega_i$ are sufficiently small such that, for each $i$, $\Omega_i$  contains at most one robot and   $ \int_{\Omega_i}\rho(z) \, dz \leq 1/N$ holds. Then $\mu\rightarrow 2$ as $|\Omega_i|\rightarrow 0$. 
\end{proposition}
This proposition and its proof appear in \cite{ICINCO2018} as Proposition 2.4.

Note that the shape of each region is also a choice that will affect the calculated value of $\mu$. Although our  approach  also requires the choice of some size and shape (namely, $\delta$ and $K$), these parameters  have much more immediate physical interpretations, making appropriate choices easier to make.

%\subsubsection{Error Metric Discretization and Desensitization}
The second, and perhaps more significant, drawback, is that by discretizing the domain we also discretize the range of values that the  error metric can assume. This means that we have simultaneously desensitized the error metric to changes in robot distribution within each region. That is, so long as the number of robots $N_i$ within each region $\Omega_i$ does not change, the distribution of robots within any and all $\Omega_i$ may be changed arbitrarily without affecting the value of $\mu$. On the other hand, the error metric $e_N^\delta$ is continuously sensitive to differences in distribution.

\subsection{Setup for Main Examples}
\label{sec:two ex}
We will perform computations with the following two desired distributions. We have purposefully chosen one that is piecewise continuous, $\rho_{\text{ring}}$, and one that is smooth, $\rho_{\text{ripple}}$.

\begin{definition}
\label{defn:ring}
The \emph{ring distribution}   $\rho_{\emph{ring}}$ is defined on the Cartesian plane with coordinates $z=(z_1$, $z_2)$ as follows. Let inner radius $r_1 = 11.4$in, outer radius $r_2 = 20.6$in, width $w = 48$in, height $h=70$in,  domain $\Omega = \{z: z_1 \in [0, w], z_2 \in [0, h]\}$, and region $\Gamma = \{z:r^2_1 < (z_1 - \frac{w}{2})^2 + (z_2 - \frac{h}{2})^2 < r^2_2\}$. Define the non-normalized $\rho'_{\emph{ring}}(z)$ to be $36 \text{ if } z \in \Omega \cap \Gamma$ and $1 \text{ if } z \in \Omega \setminus \Gamma$. Then let $C=\int_\Omega \rho'_{\emph{ring}}(z) \, dz$ and $\rho_{\emph{ring}}(z)  = C^{-1}\rho'_{\emph{ring}}(z)$ for $z\in \Omega$.
\end{definition}

\begin{definition}
\label{defn:cts}
The \emph{ripple distribution} $\rho_{\emph{ripple}}$ is defined on the Cartesian plane with coordinates $z=(z_1$, $z_2)$ as follows. Let width $w = 48$in, height $h=70$in, domain $\Omega = \{z: z_1 \in [0, w], z_2 \in [0, h]\}$, non-normalized $\rho'_{\emph{ripple}}(z) = 2+ \sin [ 3 \pi (z_1^2+z_2^2)^\frac{1}{2}] + 2 \frac{z_1^2}{w^2} + \frac{z_2^3}{h^3}$, and normalization factor $C = \int_\Omega{\rho'_{\emph{ripple}}(z)}\, dz$. Then $\rho_{\emph{ripple}}(z) = C^{-1} \rho'_{\emph{ripple}}(z)$ for $z \in \Omega$.

\end{definition}

\section{\uppercase{Error Metric Extrema}}
\label{sec:extrema}

%
%\noindent The error metric $e_N^\delta$ defined in Equation \ref{eqn:instantaneous_error_metric} inherently addresses most of the issues associated with the discretization approach. However, there is still the difficulty of interpreting whether the values of $e_N^\delta$ produced by a controller in a given situation are ``good" or not. As mentioned in Section \ref{sec:error_properties}, it is simply not possible to achieve $e_N^\delta = 0$ for every combination of target distribution $\rho$, number of robots $N$, and blob size $\delta$ . Therefore, we would like to compare the achieved value of $e_N^\delta$ against its \emph{realizable} extrema given $\rho$, $N$, and $\delta$. But $e_N^\delta$ is a highly nonlinear function of the robot positions $(x_1, ..., x_N)$,
%% As noted above, $e_N^\delta$ is a function of the robot positions $(x_1, ..., x_N)$. 
%% We are interested in finding or approximating its extrema. 
%% However, it is a highly nonlinear function, 
%and trying to find its extrema analytically has been intractable. Thus, we approach this problem by using nonlinear programming. 

\noindent In the rest of the paper, we provide tools for determining whether or not the values of $e_N^\delta$ produced by a controller in a given situation are ``good''. As mentioned in Section \ref{sec:error_properties}, it is not possible to achieve $e_N^\delta = 0$ for every combination of target distribution $\rho$, number of robots $N$, and blob radius $\delta$ . Therefore, we would like to compare the achieved value of $e_N^\delta$ against its \emph{realizable} extrema given $\rho$, $N$, and $\delta$. Unfortunately, $e_N^\delta$ is a nonlinear and moreover non-convex function of the robot positions $(x_1, ..., x_N)$,
% As noted above, $e_N^\delta$ is a function of the robot positions $(x_1, ..., x_N)$. 
% We are interested in finding or approximating its extrema. 
% However, it is a highly nonlinear function, 
and thus its extrema may elude analytical expression. Thus, we approach this problem with nonlinear programming. 

\subsection{Extrema Bounds via Nonlinear Programming}
\label{sec:extrema_bounds_calculation}
% To bound the global minimum, we first define the nonlinear optimization problem of interest. We focus on $\Omega\subset \RR^2$. 
Let $x=(x_1,..., x_N)$ represent a vector of $N$ robot coordinates. The optimization problem, first introduced in the authors' \cite{ICINCO2018}, is
\begin{align}
\label{Optimization Problem}
	&\text{minimize } e_N^{\delta}(x_1,..., x_N),   \\
	&\text{subject to } x_i \in \Omega \text{ for } i \in \{1,2,\hdots,N\}. \notag%\label{Optimization Problem} 
	%&x = (x_1,x_2,\hdots ,x_N). \notag
\end{align}
Note that the same problem structure can be used to find the maximum of the error metric by minimizing $-e_N^{\delta}$. Given $\rho$, $N$, and $\delta$, we approach these problems using a standard nonlinear programming solver, MATLAB's \texttt{fmincon}. 

A limitation of all general nonlinear programming algorithms is that successful termination produces only a local minimum, which is not guaranteed to be the global minimum. Since there is no obvious formulation of this problem for which a global solution is guaranteed,  we use the local minimum as an upper bound for the global minimum of the error metric. Heuristics such as multi-start (running the optimization many times from several initial guesses and taking the minimum of the local minima) can be used to make this bound tighter. We use $e^-$ to denote the best local minimum found and $e^+$ to denote the best local maximum found. The bounds $e^-$ and  $e^+$ serve as benchmarks against which we can compare a value of the error metric achieved by a given control law. This is reasonable, because if a configuration of robots with a lower value of the error metric exists but eludes numerical optimization, it is probably not a fair standard against which to compare the performance of a general controller.

\begin{figure}[ht]
\begin{center}
\includegraphics[width=.75\linewidth]{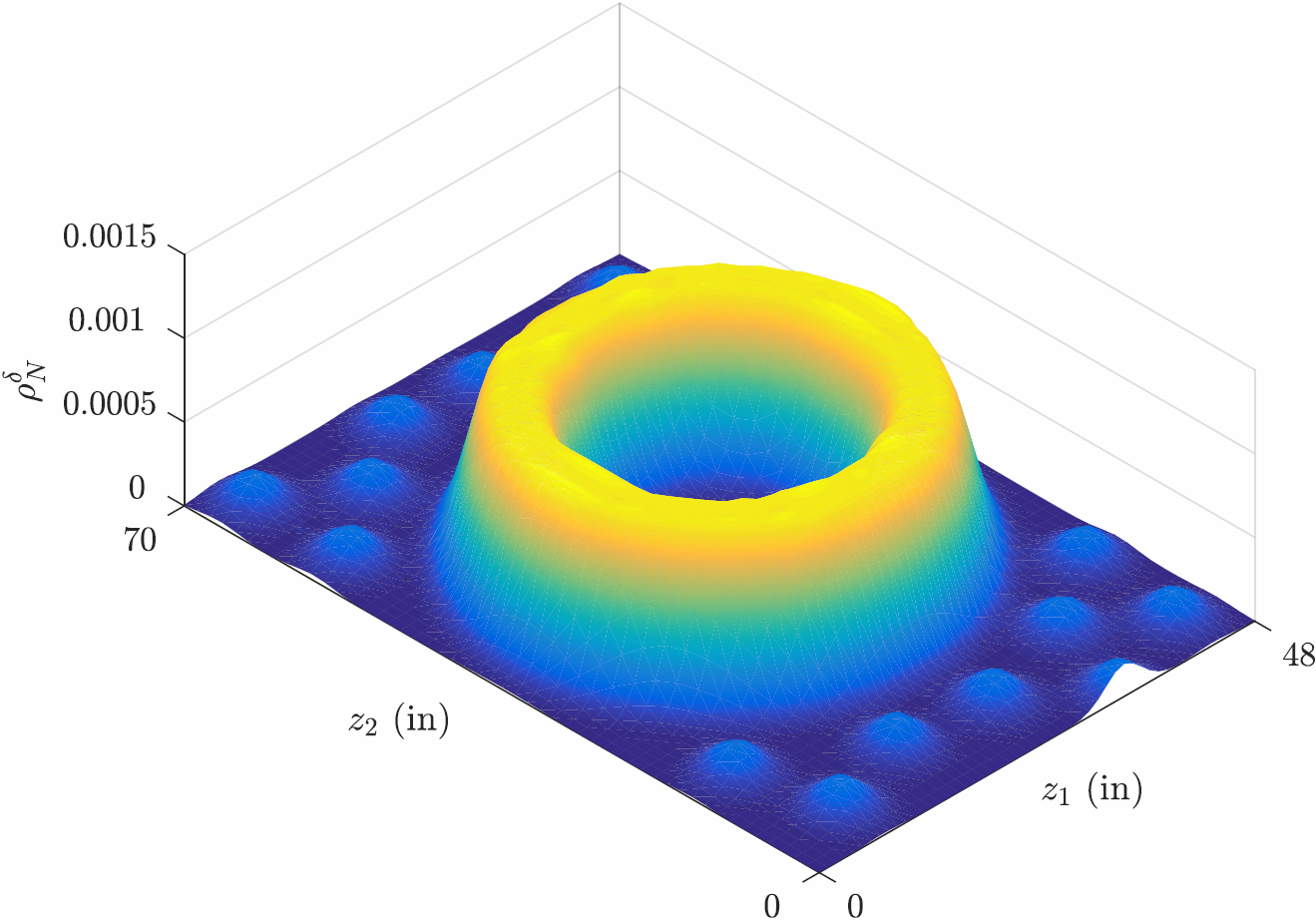}
\caption{Swarm blob function $\rho_{N=200}^{\delta = 2 \text{in}}$ corresponding with the robot distribution that yields a locally minimal value of the error metric for the ring distribution, 0.28205. This figure appears in \cite{ICINCO2018} as Figure 1.}
\label{fig:error_metric_minimum}
\end{center}
\end{figure}

\begin{figure}[ht]
\begin{center}
\includegraphics[width=.75\linewidth]{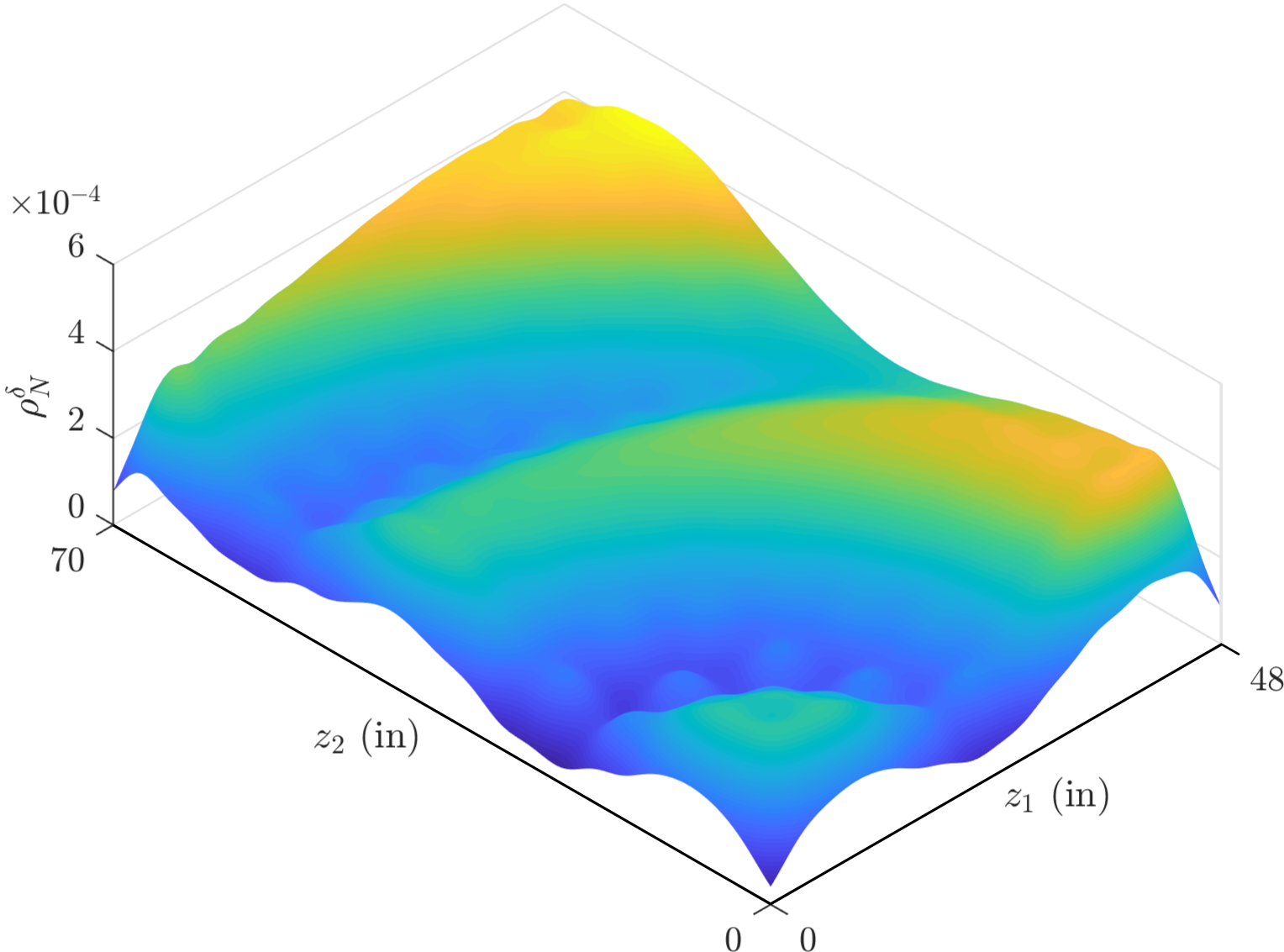}
\caption{Swarm blob function $\rho_{N=178}^{\delta = 2.5 \text{in}}$ corresponding with the robot distribution that yields a locally minimal value of the error metric for the ripple distribution, 0.06376.}
\label{fig:ripple_error_metric_maximum}
\end{center}
\end{figure}

\begin{figure}[ht]
\begin{center}
\includegraphics[width=.75 \linewidth]{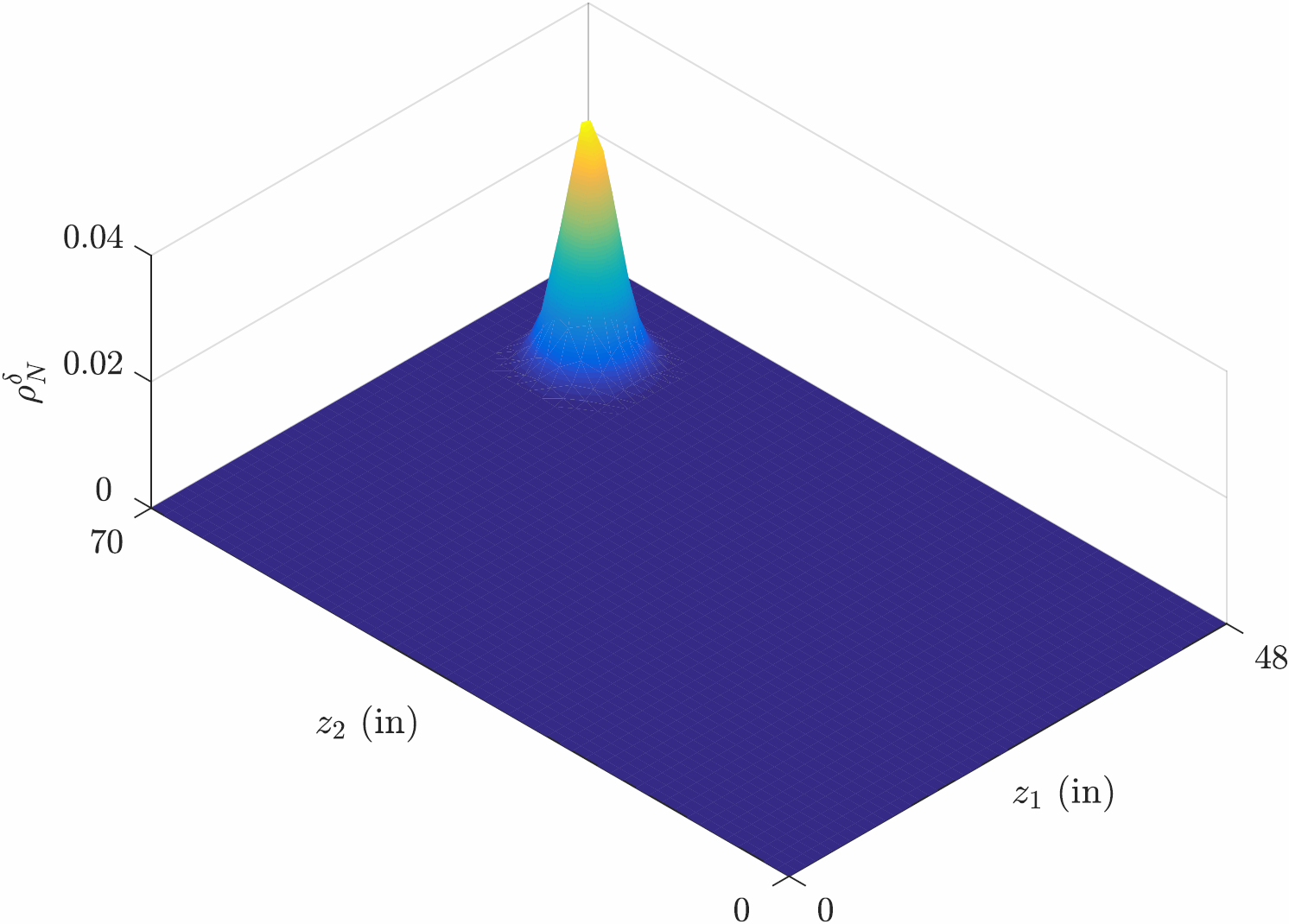}
\caption{Swarm blob function $\rho_{N=200}^{\delta = 2 \text{in}}$ corresponding with the robot distribution that yields a locally maximal value of the error metric for the ring distribution, 1.9867. This occurs when all robots coincide outside the ring.  This figure appears in \cite{ICINCO2018} as Figure 2.}
\label{fig:error_metric_maximum}
\end{center}
\end{figure}

Examples of extremal swarm blob functions found using this approach are shown in Figures \ref{fig:error_metric_minimum}, \ref{fig:ripple_error_metric_maximum}, and \ref{fig:error_metric_maximum}.

\subsubsection{Relative Error}
We introduce the notion of \emph{relative error}, a quantitative way of assessing the performance of a robot distribution controller. This notion first appeared in the authors' \cite{ICINCO2018}. 
\begin{definition}
\label{defn:erel}
Let $e_{\emph{observed}}$ denote the error value of a robot configuration. 
The \emph{relative error} is defined as,
\begin{equation}
\label{eqn:erel}
e_\emph{rel} = \frac{e_{\emph{observed}} - e^-}{e^+ - e^-}.
\end{equation}
\end{definition}
In order to apply this definition, we need to specify how to find $e_{\text{observed}}$. If the robot positions $x_1,..., x_N$ produced by a given controller are constant, then $e_{\text{observed}}$ can simply be taken as $e_N^\delta(x_1,..., x_N)$. In general, however, the positions  $x_1,..., x_N$ may change over time, and it is natural to define $e_{\text{observed}}$ based on the steady state value of $e_N^\delta$. We suggest defining the \emph{steady state settling time} $t_s$ to be the time at which the error has settled to within $2\%$ of its asymptotic value. Then, we propose taking $e_\text{observed}$ to be the third-quartile value observed for $t > t_s$, which we denote $e_\text{Q3}$.

%The performance of a robot distribution controller can be quantitatively assessed by calculating the error value $e_{\text{observed}}$ of a robot configuration it produces, and comparing this value against the extrema bounds $e^-$ and $e^+$. If the robot positions $x_1,..., x_N$ produced by a given controller are constant, then $e_{\text{observed}}$ can simply be taken as $e_N^\delta(x_1,..., x_N)$. In general, however, the positions  $x_1,..., x_N$ may change over time. In this case, we suggest using the third-quartile value observed after the system reaches steady state, which we denote $e_\text{Q3}$.

%Consider the relative error
%\begin{equation*}
%\label{eqn:erel}
%e_\text{rel} = \frac{e_{\text{observed}} - e^-}{e^+ - e^-}.
%\end{equation*}
We suggest that if $e_\text{rel}$ is less than 10\%, the performance of the controller is quite close to the best possible, and if this ratio is 30\% or higher, the performance of the controller is rather poor.

We summarize the procedure described in this section:
\begin{itemize}
\item Step 1: Given $\rho$, $N$ and $\delta$, compute $e^-$ and $e^+$. 
\item Step 2: Compute $e_{\text{observed}}$ for the desired swarm controller. That is, 
\begin{itemize}
\item Step 2.a: Calculate $e_N^\delta(t)$ from robot trajectories $x_1(t), ..., x_N(t)$ produced by the controller.
\item Step 2.b: Find the steady state settling time $t_s$.
\item Step 2.c: Measure the third quartile value of the error metric $e_{Q_3}$ for $t>t_s$. This is  $e_{\text{observed}}$.
\end{itemize}
\item Step 3: Evaluate $e_{\text{rel}}$ according to Definition \ref{defn:erel}.
\end{itemize}
We emphasize that $e^-$ and $e^+$ are independent of the particular controller; therefore, Step 1 has to be completed only once when comparing different controllers or different outcomes of running one controller. %, so long as the desired distribution $\rho$ remains the same.

%\subsection{Relative Error}
%The performance of a robot distribution controller can be quantitatively assessed by calculating the error value $e_{\text{observed}}$ of a robot configuration it produces, and comparing the value against the extrema bounds $e^-$ and $e^+$. We define the relative error as,
%\begin{equation*}
%%\label{eqn:erel}
%e_\text{rel} = \frac{e_{\text{observed}} - e^-}{e^+ - e^-},
%\end{equation*}
%where $e_{\text{observed}}$ can be obtained from $e_N^\delta$ in the following two ways.
%
%If a given controller outputs final robot positions $x_1,..., x_N$, then, to measure the performance of the controller, $e_{\text{observed}}$ can simply be taken to be $e_N^\delta(x_1,..., x_N)$. 
%
%On the other hand, if the positions  $x_1,..., x_N$ continue to change over time, we suggest using the third-quartile value observed after the system reaches steady state. We will denote this by $e_\text{Q3}$, so that the relative error becomes,
%\begin{equation}
%\label{eqn:erel}
%e_\text{rel} = \frac{e_\text{Q3} - e^-}{e^+ - e^-}.
%\end{equation}
%
%In either situation, we suggest that if $e_\text{rel}$ is less than 10\%, the performance of the controller is quite close to the best possible, whereas if this ratio is 30\% or higher, the performance of the controller is rather poor. 

\subsubsection{Example}
\label{example optimization}
We apply this method to assess the performance of the controller in \cite{li2017decentralized}, which guides a swarm of $N=200$ robots with $\delta = 2$in (the physical radius of the robots) to achieve the ring distribution from Definition \ref{defn:ring}. These calculations were originally performed in the authors' \cite{ICINCO2018}.

\noindent\textbf{Step 1: Compute $e^-$ and $e^+$. }
To determine an upper bound on the global minimum of the error metric, we computed 50 local minima of the error metric starting with random initial guesses, then took the lowest of these to be $e^- = 0.28205$. An equivalent procedure bounds the global maximum as $e^+ = 1.9867$, produced when all robot positions coincide near a corner of the domain. The corresponding swarm blob functions are depicted in Figures \ref{fig:error_metric_minimum} and \ref{fig:error_metric_maximum}. 
% These results are shown in Figure \ref{fig:error_metric_minimum} and Figure \ref{fig:error_metric_maximum}. 
Note that the minimum of the error metric is significantly higher than zero for this finite number of robots of nonzero radius, emphasizing the importance of performing this benchmark calculation rather than using zero as a reference.

\noindent\textbf{Step 2: Find $e_{\text{observed}}$.} Under the   stochastic control law of \cite{li2017decentralized}, the behavior of the error metric over time appears to be a noisy decaying exponential. Therefore, we fit to the data shown in Figure 7 of \cite{li2017decentralized} a function of the form $f(t) = \alpha + \beta \exp(-\frac{t}{\tau})$ by finding error asymptote $\alpha$, error range $\beta$, and time constant $\tau$ that minimize the sum of squares of residuals between $f(t)$ and the data. By convention, the steady state settling time is taken to be $t_\text{s} = 4 \tau$, which can be interpreted as the time at which the error has settled to within $2\%$ of its asymptotic value \cite{barbosa2004tuning}. The third quartile value of the error metric for $t > t_\text{s}$ is $e_{Q3} = 0.5157$.

\noindent\textbf{Step 3: Find $e_{\text{rel}}$.} Using these values for $e_\text{Q3}$, $e^-$, and $e^+$, we calculate $e_\text{rel}$ according to Equation \ref{eqn:erel} as $13.71\%$.

\begin{remark}
Although the sentiment of the $e_\text{rel}$ benchmark is found in \cite{li2017decentralized}, we have formalized the calculation and made three important improvements to make it suitable for general use. First, we have placed the calculation of  $e^-$ and $e^+$ into a framework appropriate for nonlinear programming, so that the calculation is objective and repeatable. On the other hand, in \cite{li2017decentralized}, values analogous to  $e^-$ and $e^+$ were found by ``manual placement'' of robots. Second, we suggest a general way of evaluating error in time-dependent controllers. In particular, although  \cite{li2017decentralized} refers to steady state, a  definition is not suggested there. Adopting the 2\% settling time convention allows for an unambiguous calculation of $e_{Q_3}$ and other steady state error metric statistics. It also provides a metric for assessing the speed with which the control law effects the desired distribution.  Finally,  we suggest using the third quartile value $e_{Q_3}$ in the calculation, in contrast to the minimum observed value of the error metric used in \cite{li2017decentralized}. Since $e_{Q_3}$ better represents the distribution of error metric values achieved by a controller, our notion of relative error is more representative of the controller's overall performance.

%\cite{li2017decentralized} uses the \emph{minimum observed value} of the error metric in the calculation, but we suggest that the third quartile value better represents the distribution of error metric values achieved by a controller, and thus is more representative of the controller's overall performance. 

These changes account for the difference between our calculated value of $e_\text{rel} = 13.71\%$ and the report in \cite{li2017decentralized} that the error is  ``$7.2$\% of the range between the minimum error value\dots{} and maximum error value''. Our substantially higher value of $e_\text{rel}$ indicates that the performance of this controller is not very close the best possible. We emphasize this to motivate the need for our second benchmark in Section \ref{sec:errorpdf}, which may be more appropriate for a stochastic controller like the one presented in \cite{li2017decentralized}.
\end{remark}
\subsection{Error Metric for Optimal Swarm Design}
\label{subsec:swarm design}
\begin{figure*}
\includegraphics[width=\linewidth]{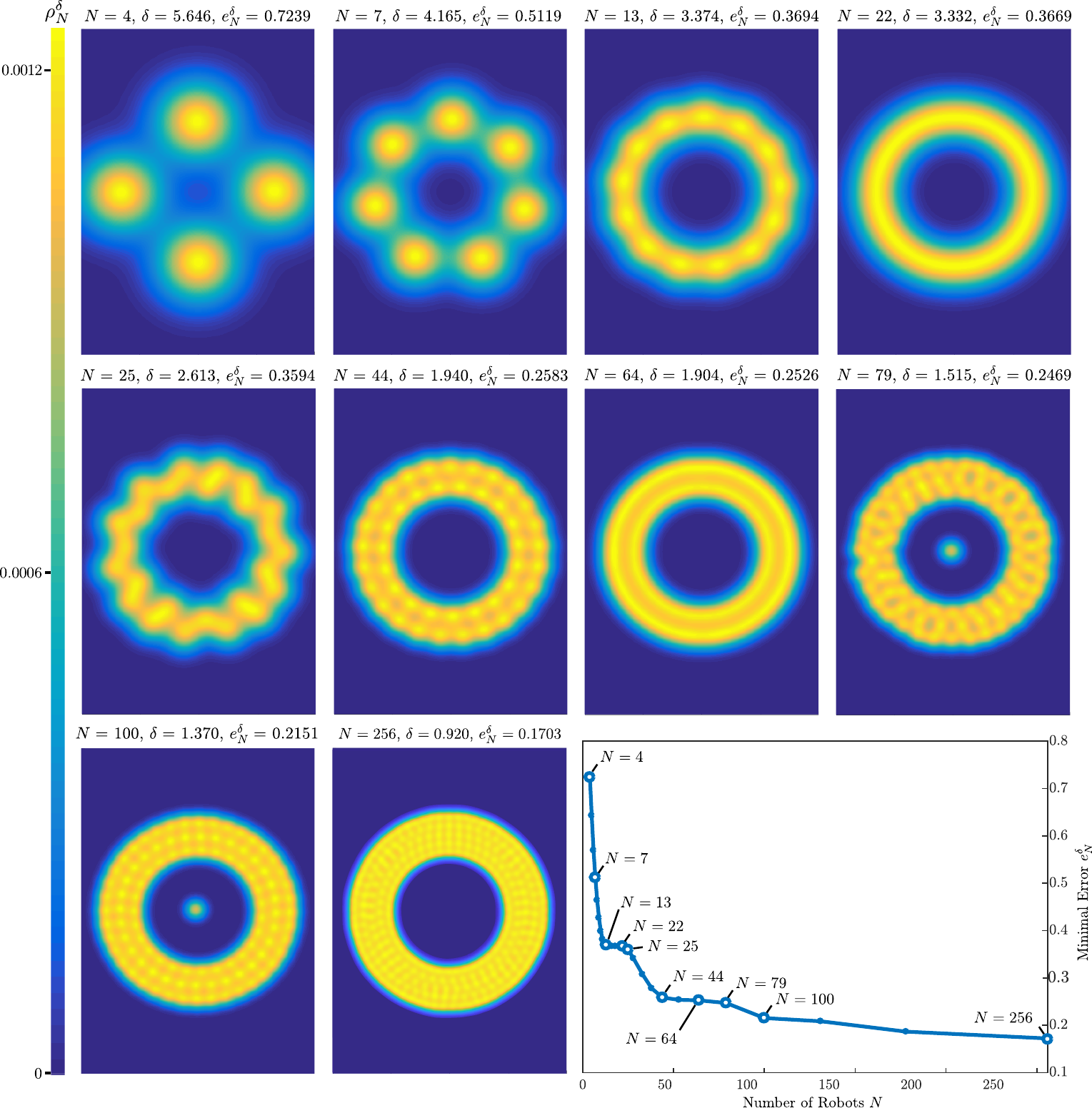}
\centering
\caption{Swarm blob functions $\rho_{N}^{\delta}$ corresponding with the robot distributions and values of $\delta$ that yield the minimum value of the error metric for the ring distribution target. These plots appear in Figure 3 of \cite{ICINCO2018}. Inset graph shows the relationship between $N$ and the minimum value of the error metric observed from repeated numerical optimization. For $N < 256$, the initial guess provided to the optimizer had robots uniformly randomly distributed within the domain. For $N=256$ in \cite{ICINCO2018}, such a guess resulted in local minima with error values \emph{greater} than those for smaller swarms, inconsistent with the intuitive notion that a larger swarm should be able to more accurately achieve the target distribution. To remedy this here, the initial guesses for $N=256$ and higher (not shown) were taken with all robots between the inner and outer radius of the ring (i.e. within the region $\Gamma$ from Definition \ref{defn:ring}). With these the plot decreases monotonically for all values of $N$ tested (including 300, 350, 400, 450, and 500), but progress appears slow beyond $N=256$; doubling the number of robots decreases the error metric by only 10\%.
}
\label{fig:blobs}
\end{figure*}

\begin{figure*}[t]
\includegraphics[width=\linewidth]{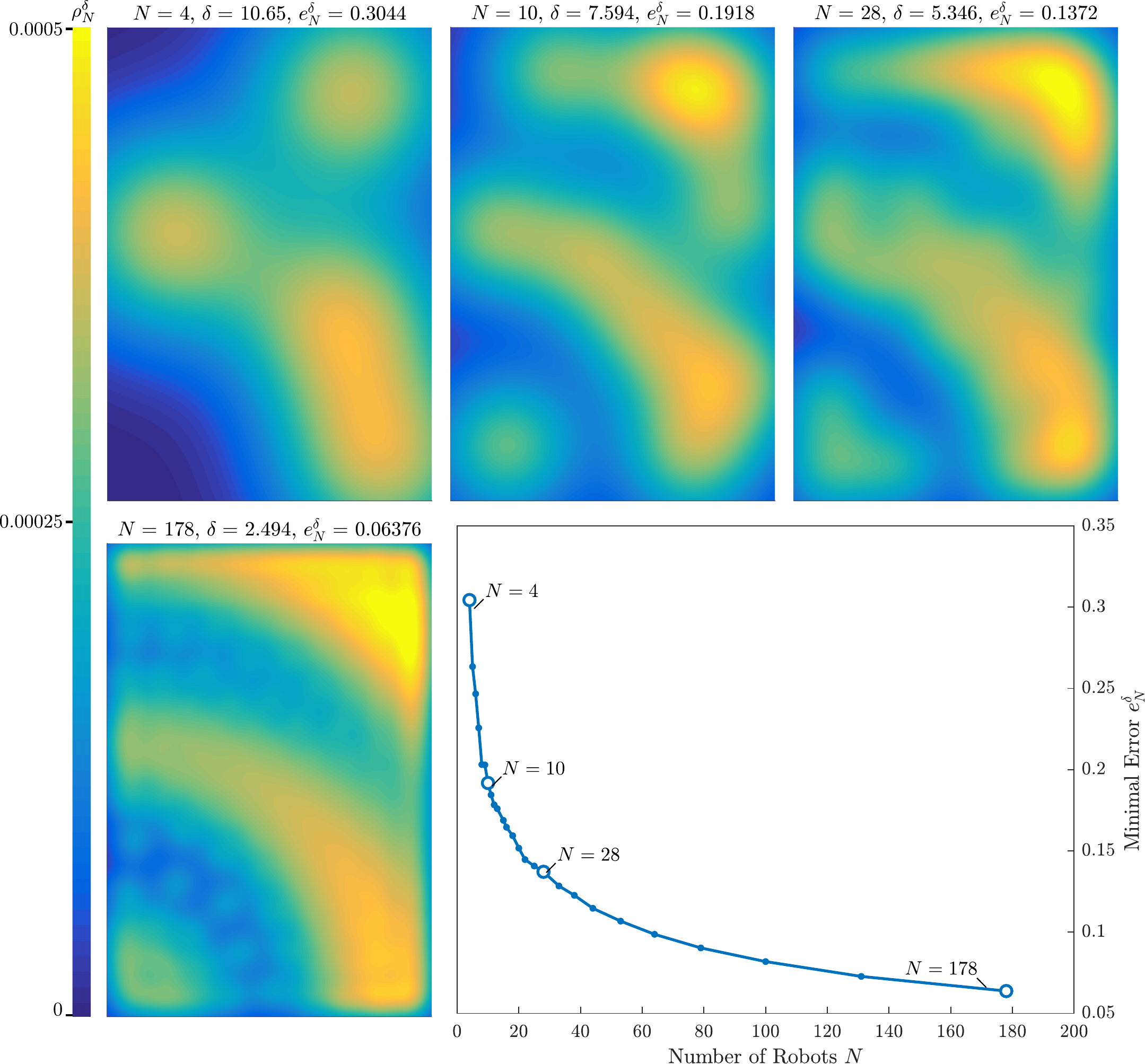}
\centering
\caption{Swarm blob functions $\rho_{N}^{\delta}$ corresponding with the robot distributions and values of $\delta$ that yield the minimum value of the error metric for the ripple distribution target. Inset graph shows the relationship between $N$ and the minimum value of the error metric observed from repeated numerical optimization.}
\label{fig:blobs2}
\end{figure*}

\begin{figure}[h]
\begin{center}
\includegraphics[width=.75\linewidth]{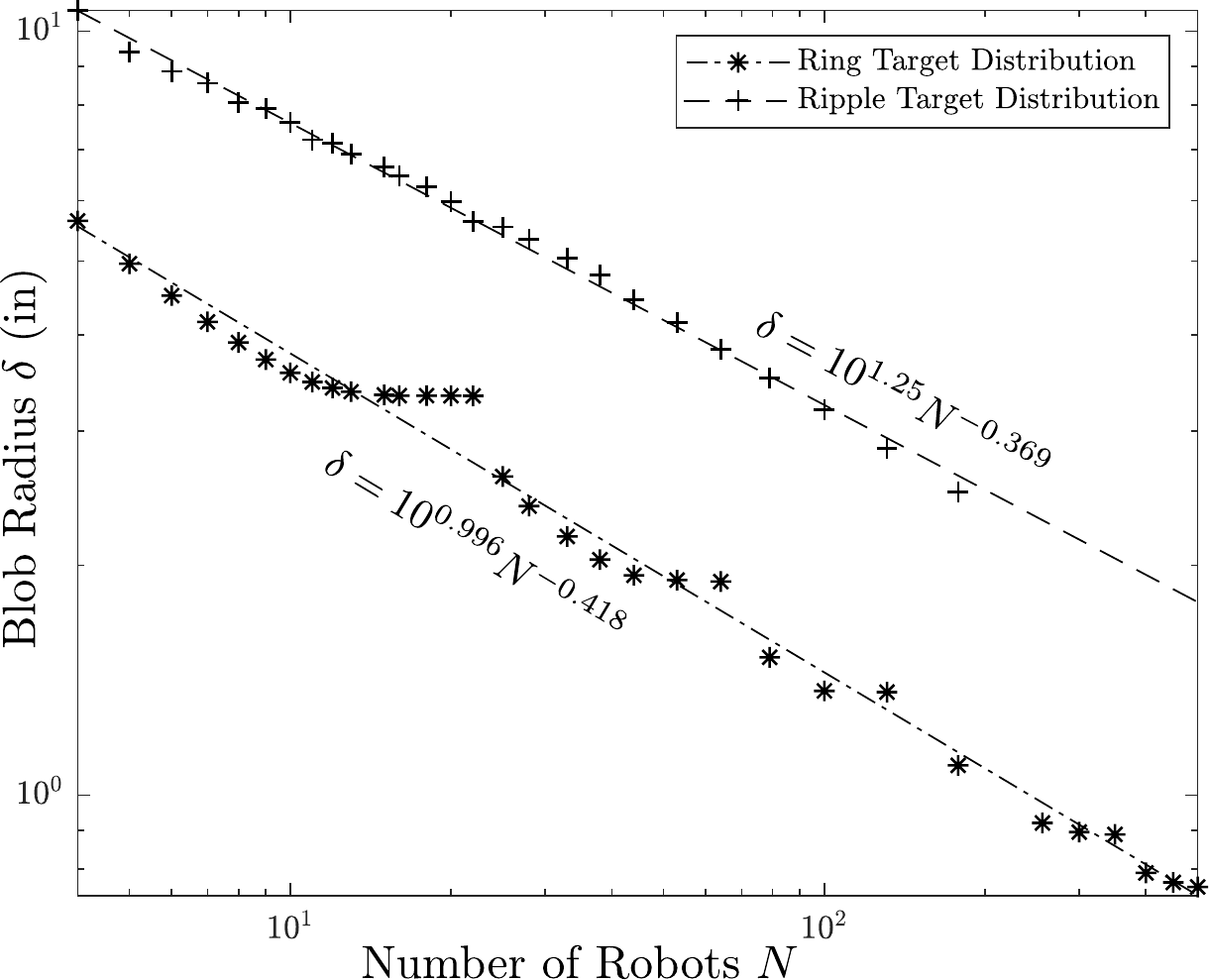}
\caption{Optimal blob radius $\delta$ to minimize error $e_N^\delta$ along with best fit lines ($\log$-$\log$ scale). Note that for both target distributions the optimal blob radius seems to scale with $N^p$ where $p \approx -0.4$. This relationship may be useful for approximating the effective radius of robot sensing/manipulation capabilities in order to maximize effectiveness for a given swarm size.}
\label{fig:deltaVN}
\end{center}
\end{figure}

So far we have taken $N$ and $\delta$ to be fixed; we have assumed that the robotic agents and size of the swarm have already been chosen. Now, we briefly consider the use of the error metric as an objective function for the \emph{design} of a swarm. 
% \subsubsection{Minimizing over values of the radius $\delta$}
Adding $\delta > 0$ as a decision variable to (\ref{Optimization Problem}) yields the minimization problem,
\begin{align}
\label{Optimization Problem delta}
	&\text{minimize } e_N^{\delta}(x_1,..., x_N),   \\
	&\text{subject to } x_i \in \Omega \text{ for } i \in \{1,2,\hdots,N\}, \, \delta>0. \notag%\label{Optimization Problem} 
	%&x = (x_1,x_2,\hdots ,x_N). \notag
\end{align}
Solving (\ref{Optimization Problem delta}) for several fixed values of $N$ provides insight into the number of robots and what effective working radius are needed to achieve a given level of coverage for a particular target distribution. Visualizations of such calculations are provided in Figure \ref{fig:blobs} and \ref{fig:blobs2}, and Figure \ref{fig:deltaVN} shows the optimal value of $\delta$ as a function of swarm size.

Note that ``breakthroughs'', or relatively rapid decreases in the error metric, can occur once a critical number of robots is available; these correspond with a qualitative change in the distribution of robots. For example, at $N=22$ in Figure \ref{fig:blobs} the robots are arranged in a single ring; beginning with $N=25$ we see the robots start to be arranged in two separate concentric rings of different radii and the error metric begins to drop sharply. On a related note, there are also ``lulls'' in which increasing the number of robots has little effect on the minimum value of the error metric, such as between $N=44$ and $N=79$. We leave for future work the task of finding a theoretical explanation for these breakthroughs and lulls. In the meantime, computations like these can help a swarm designer determine the best number of robots $N$ and effective radius of each $\delta$ to achieve the required coverage.

%\subsubsection{The radius $\delta$}
%\label{subsec:minimizing delta}
%We also computed what value of $\delta>0$ corresponds to the minimum of (\ref{Optimization Problem delta}) for a given $N$. We use $\delta(N)$ to refer to this minimizing $\delta$. We find that $\delta(N)\approx ??????????????$ when the desired distribution is $\rho_{\text{ripple}}$ and $\delta(N)\approx ?????????????$ when the desired distribution is $\rho_{\text{ring}}$.

\section{\uppercase{Error Metric Probability Density Function}}
\label{sec:errorpdf}
\noindent In the previous section, we  described how to find bounds on the minimum and maximum values for error and use these as benchmarks against which to measure the performance of a given control law. However, stochastic control laws may tend to produce robot positions with observed error $e_\text{observed}$ well above the minimum $e^-$. For these controllers, $e_\text{rel}$ may be too stringent of a measure for assessing whether the controller's performance is ``good'' or ``bad'', and so we propose an alternative. %A question that remains is, how ``easy" or ``difficult" is it to achieve such  values? 
%This is important in order to use the error metric to assess the effectiveness of an underlying control law. 

According to the setup of our problem, the goal of any  control law is for the robots to achieve the desired distribution $\rho$. Thus, it is natural to compare the outcome of a given control law to simply picking the robot positions at random from the target distribution $\rho$. % In this section we consider the robot positions as being sampled directly from the desired distribution and study the statistical properties of the error metric in this situation both analytically and numerically. %\textcolor{blue}{Moreover, we suggest a second benchmark} with which to evaluate the performance of a swarm distribution controller. %Specifically, we compare the location and spread of the probability density function (PDF) of the error metric for the swarm distribution controller against the error metric PDF when robot positions are sampled from the desired distribution.
%We take the robots' positions $X_1$, \dots, $X_N$, to be independent, identically distributed bivariate random vectors in $\Omega\subset \RR^2$ with probability density function $\rho$. We place a blob of shape $K$ at each of the $X_i$ (previously we took $K$ to be the Gaussian $G$), so that the swarm blob function is,
%\begin{equation}
%\label{rho random}
%\rho_N^\delta (z) = \frac{1}{N\delta^2}\sum_{i=1}^N K^\delta\left(z - X_i\right),
%\end{equation}
%where $K^\delta$ is defined by (\ref{def:Kdelta}). We point out that the right-hand side of (\ref{rho random}) is exactly that of (\ref{eqn:rho with N}) upon taking $K$ to be the Gaussian $G$ and the robot locations $x_i$ to be the randomly selected $X_i$. The error $e_N^\delta$ is now a random variable, the value of which depends on the particular realization of the robot positions $X_1$, \dots, $X_N$, but which has a well-defined PDF. The performance of a stochastic robot distribution controller can be quantitatively assessed by calculating the error values $e_N^\delta(x_1,..., x_N)$ it produces in steady state and comparing their distribution to the PDF described above. 
We take the robots' positions $X_1$, \dots, $X_N$, to be independent, identically distributed bivariate random vectors in $\Omega\subset \RR^2$ with probability density function $\rho$. We place a blob of shape $K$ at each of the $X_i$ (previously we took $K$ to be the Gaussian $G$), so that the swarm blob function is,
\begin{equation}
\label{rho random}
\rho_N^\delta (z) = \frac{1}{N}\sum_{i=1}^N K^\delta\left(z - X_i\right),
\end{equation}
where $K^\delta$ is defined by (\ref{def:Kdelta}). Equation (\ref{rho random}) appears in the authors' \cite{ICINCO2018}. We point out that the right-hand sides of (\ref{rho random}) and (\ref{eqn:rho with N}) agree upon taking $K$ to be the Gaussian $G$ in (\ref{rho random}) and the robot locations $x_i$ to be the randomly selected $X_i$ in (\ref{eqn:rho with N}).   Moreover, we have the relationship,
\begin{equation}
\label{eq:convolve}
\rho_N^\delta = K^\delta * \left(\frac{1}{N}\sum_{i=1}^N \delta_{X_i}\right),
\end{equation}
where $*$ denotes convolution. This equality is the reason (\ref{eq:delta masses}) holds.  

In this context, the error $e_N^\delta$ is a random variable, the value of which depends on the particular realization of the robot positions $X_1$, \dots, $X_N$, and thus has a well-defined probability density function (PDF) and cumulative distribution function (CDF). We denote the PDF and CDF by $f_{e_N^\delta}$ and $F_{e_N^\delta}$, respectively. The performance of a stochastic robot distribution controller can be quantitatively assessed by calculating the error values  it produces in steady state and comparing their distribution to $f_{e_N^\delta}$. We introduce this benchmark in Subsection \ref{subsec:2nd benchmark}.

\subsection{Kernel Density Estimation}
\label{sec:kde}
The approach we take in this section is closely linked to the statistical theory of kernel density estimation (KDE). For an overview, see, for example, the book \cite{Wand}. Broadly, the aim of this area of study is to find the underlying density $\rho$ from the values of identically distributed random variables sampled from this density. The expression (\ref{rho random}) is the so-called \emph{kernel density estimator} of $\rho$, and $\rho_N^\delta$ is considered as an approximation to $\rho$.  In a sense, this is the reverse point of view from the one we take, since we are given $\rho$ and seek to learn about the distribution of the $X_i$s. Nevertheless, we are able to apply ideas from this well-developed field of study to our context.

There are many notions of error that are used in the context of KDE. We refer the reader to \cite[Chapter 2]{WandJones} for a summary. Our notion $e_N^\delta$ corresponds to \emph{integrated absolute error} (IAE). The use of IAE instead of other notions, particularly ones involving the $L^2$ norm, was advocated for in \cite{DG}.

We conclude this subsection by applying two results from KDE theory to ideas we've presented so far. Then,  in Subsection \ref{theory converge} we present rigorous results that show that the error metric has an approximately normal distribution when the $X_i$ are sampled from $\rho$. As a corollary we obtain that the limit of this error is zero  as $N$ approaches infinity and $\delta$ approaches 0. Subsections \ref{subsec:num} and \ref{subsec: num ex} include a numerical demonstration of these results. 

The theoretical results presented in the next subsection not only support our numerical findings, but they also allow for faster computation. Indeed, if one did not already know that the error when robots are sampled randomly from $\rho$ has a normal distribution for large $N$, tremendous computation may be needed to get an accurate estimate of this probability density function. On the other hand, since the results we present prove that the error metric has a normal distribution for large $N$, we need only fit a Gaussian function to the results of relatively little computation. In Subsection \ref{subsec: num ex} we present an example calculation demonstrating the utility of this approach.% only a sample mean and variance need to be calculated in order to approximate the entire CDF and PDF.

% As a benchmark, we suggest that if nearly half of the configurations produced by a given controller yield an error metric value that lies within the middle quartiles of this error metric PDF, then the performance appears consistent with that expected of a stochastic swarm controller such as that of \cite{li2017decentralized}. 

\subsubsection{Optimal $\delta(N)$}
\label{subsec:opt delta}
In Subsection \ref{subsec:swarm design}, we included $\delta$ as a parameter in the optimization problem (\ref{Optimization Problem}) and numerically obtained an estimate for the optimal $\delta$ given a swarm size $N$. This problem has also been studied in the context of KDE. In \cite{ScottWand} it was found that the optimal choice is, 
%In Subsection \ref{subsec:swarm design}, we included $\delta$ as a parameter in the optimization problem (\ref{Optimization Problem}). The related problem of what is the optimal radius $\delta$ given a swarm size $N$  has  been studied in the context of KDE. In \cite{ScottWand} it was found that the optimal choice is, 
\[
\delta^*(N) = CN^{-1/6},
\] 
where $C$ is a constant that does not depend on $N$. This is equation (2.3) of \cite{ScottWand} in the 2-dimensional case. Here, ``optimal'' means that this choice of $\delta$ minimizes the expectation of the $L^1$ error between $\rho$ and $\rho_N^\delta$. This result holds under the hypotheses that $K$ satisfies (\ref{eq:Kcondition}) and $\rho$ has bounded second derivatives. 

We remark that the optimal $\delta(N)$  minimizing (\ref{Optimization Problem delta}) that was found in Subsection \ref{subsec:swarm design} and shown in Figure \ref{fig:deltaVN} scales with $N^{p}$ where $p \approx -0.4 \neq -1/6$. Despite the difference in exponent, these computations do not contradict the results of \cite{ScottWand} as the latter considers the radius $\delta$ that minimizes the \emph{expected} error value, and hence takes into account robot positions that are not optimal. On the other hand, the values of $\delta(N)$ computed in Subsection \ref{subsec:swarm design} are optimal only in concert with the optimal robot positions.  %Finding a quantitative explanation for the contrast between the two situations is an interesting direction for future work. 
The contrast between the two situations suggests that for the design of a swarm as described in Subsection \ref{subsec:swarm design}, the sensing/manipulation radius required to maximize coverage may depend on whether the controller is stochastic or deterministic.

%HERE WILL BE A DISCUSSION COMPARING THIS EXPONENT TO WHAT WAS FOUND IN SUBSECTION \ref{subsec:swarm design}.

\subsubsection{Choice of $K$}
\label{subsec:K}
In the context of KDE, there is a notion of how efficient a given kernel $K$ is. This notion is formulated in terms of $L^2$ error between $\rho$ and $\rho_N^\delta$, and therefore the precise details do not directly apply to our situation. However, we point out that it is known \cite[Section 2.7]{WandJones} that, so long as the kernel $K$ is non-negative, has integral 1, and satisfies
\begin{equation}
\label{eq:Kcondition}
\int zK(z)\, dz=0, \quad \int z^2 K(z)\, dz<\infty,
\end{equation}
then the particular choice of $K$ has only a marginal effect on efficiency. We leave to future work the extension of this concept to our context.

%There are several ways to construct multivariate kernels from single variable functions

\subsection{Theoretical Central Limit Theorem}
\label{theory converge}
It turns out that, under appropriate hypotheses, the $L^1$ error between $\rho$ and $\rho^\delta_N$ has a normal distribution with mean and variance that approach zero as $N$ approaches infinity.  In other words, a central limit theorem holds for the error. The first such result was obtained in \cite{DG} in the one-dimensional case. Previously, similar results, such as those in \cite{BR}, were available only for $L^2$ notions of error, which are easier to analyze.

For  any result of central limit theorem type to hold, $\delta$ and $N$ have to be compatible. Thus, for the remainder of this subsection $\delta$ will depend on $N$, and we display this as $\delta(N)$. 
We have,
\begin{theorem}
\label{thm:H}
Suppose $\rho$ is  twice continuously differentiable, and $K$ is zero outside of some bounded region and radially symmetric. Then,  for $\delta(N)$ satisfying
\begin{equation}
\label{condition delta}
\delta(N)=O(N^{-1/6}) \text{ and } \lim_{N\rightarrow \infty}\delta(N) N^{1/4}=\infty,
\end{equation}
we have 
\[
e^{\delta(N)}_N \approx \mathcal{N}\left(\frac{e(N)}{N^{1/2}}, \frac{\sigma(N)^2\delta(N)^2}{N}\right),
\]
where $\sigma^2(N)$ and $e(N)$ are deterministic quantities that are bounded uniformly in $N$.\footnote{Here $\mathcal{N}(\mu, \sigma^2)$ denotes the normal random variable of mean $\mu$ and variance $\sigma^2$, and we use the notation $\approx$ to mean  that the difference of the quantity on the left-hand side and on the right-hand side converges to zero in the sense of distributions as $N\rightarrow\infty$.}
\end{theorem}

From Theorem \ref{thm:H} it is easy to deduce:
\begin{corollary}
\label{cor:H}
Under the hypotheses of Theorem \ref{thm:H}, the error $e_N^{\delta(N)}$ converges in distribution to zero as $N\rightarrow \infty$.
\end{corollary}

Corollary \ref{cor:H}, Theorem \ref{thm:H}, and the proof of Theorem \ref{thm:H} (which follows from \cite[Theorem, page 1935]{horvath1991lp}) appear in the authors' \cite{ICINCO2018}.

\begin{remark}
\label{rem:smooth} 
There are a few ways in which practical situations may not align perfectly with the assumptions of Theorem \ref{thm:H}. However, we posit that in all of these cases, the differences are numerically insignificant. Moreover, we believe that a stronger version of Theorem \ref{thm:H} may hold, one which applies in our situation. We now briefly summarize these three discrepancies and indicate how to resolve them. 

First, we defined our density $\rho_N^\delta$ by (\ref{eqn:normalized_blob_function}), but in this section we use a version with denominator $N$. However, as explained above, the two expressions approach each other for small $\delta$, and this is the situation we are interested in here. We believe that the result of Theorem \ref{thm:H} should hold with (\ref{eqn:normalized_blob_function}) instead of (\ref{rho random}).

Second, in one of the two examples we consider here, the desired density $\rho$ is only piecewise continuous but  not twice differentiable. We point out that an arbitrary density $\rho$ may be approximated to arbitrary precision by a smoothed out version, for example by convolution with a mollifier (a standard reference is \cite[Section 4.4]{brezis2010functional}). Moreover, the main results of \cite{GMZ} give a version of Theorem \ref{thm:H} that require that $\rho$ is only Lebesgue measurable (piecewise continuous functions satisfy this hypothesis). However, the  results of \cite{GMZ} only apply in 1 dimension. Generalizing the results of \cite{GMZ} to several spatial dimensions is an interesting open problem in KDE, but is beyond the scope of our work here.

 Third, in our computations we use the kernel $G$, which is not compactly supported, for the sake of simplicity. Similarly, this kernel can be approximated, with arbitrary accuracy, by a compactly supported version. Making these changes to the kernel or target density would not affect the conclusions of numerical results.
\end{remark}

\subsection{Numerical Approximation of the Error Metric PDF}
\label{subsec:num}
%In this subsection we describe how to numerically find the CDF and PDF of $e_N^\delta$ when the robot positions are randomly sampled from $\rho$.  
In this subsection we describe how to numerically find $f_{e_N^\delta}$ and $F_{e_N^\delta}$.  
According to Theorem \ref{thm:H}, for sufficiently large $N$, one could simply use random sampling to estimate the mean and standard deviation, then take these as the parameters of the normal PDF and CDF. However, for moderate $N$, we choose to begin by estimating the entire CDF and confirming that it is approximately normal.  To this end, we apply:
\begin{proposition}
\label{prop:cdf} We have, %Let  $F_{e_N^\delta}$ denote the CDF of $e_N^\delta$ when the robot positions are randomly sampled from $\rho$. Then,
\begin{equation}
F_{e_N^\delta}(z)=\int_{\Omega^N} \boldsymbol{1}_{\{ x\mid e_N^{\delta}(x)\le z \}} \prod^N_{i=1} \rho( x_i)  dx. \label{Derived Distribution}
\end{equation}
\end{proposition}
Here $\boldsymbol{1}$ denotes the indicator function. Proposition \ref{prop:cdf} and its proof appear in the authors' \cite{ICINCO2018}.

Notice that, since each of the $x_i$ is itself a 2-dimensional vector as the $X_i$ are random points in the plane,
% and we are using the notation $x=(x_1,..., x_N)$,  
the integral defining the cumulative distribution function of the error metric is of dimension $2N$. Finding analytical representations for the CDF %is combinatorially complex and
quickly becomes infeasible for large swarms. Therefore, we approximate  (\ref{Derived Distribution}) using Monte Carlo integration, which is well-suited for high-dimensional integration \cite{sloan2010integration}\footnote{Quasi-Monte Carlo techniques, which use a low-discrepancy sequence rather than truly random evaluation points, promise somewhat faster convergence but require considerably greater effort to implement. The difficulty is in generating a low-discrepancy sequence from the desired distribution, which is possible using the Hlawka-M\"uck method, but computationally expensive \cite{hartinger2006non}.}. Next, we fit a Gauss error function ($\erf(\cdot)$, the integral of a Gaussian $G$) to the data. If the fitted curve matches the data well, we differentiate to obtain the PDF. We remark that we express the integral in (\ref{Derived Distribution}) in terms of an indicator function in order to express the quantity of interest in a way that is easily approximated with Monte Carlo integration.

\subsubsection{Computation}
\label{subsec:compute f F}
We apply Proposition \ref{prop:cdf} to numerically find the PDF $f_{e^\delta_N}$ and the CDF $F_{e^\delta_N}$  for $\rho = \rho_{\text{ring}}$ (see Definition \ref{defn:ring}),  $N=200$, and $\delta=2\text{in}$.

\begin{figure}[ht]
\begin{center}
\includegraphics[width=.75\linewidth]{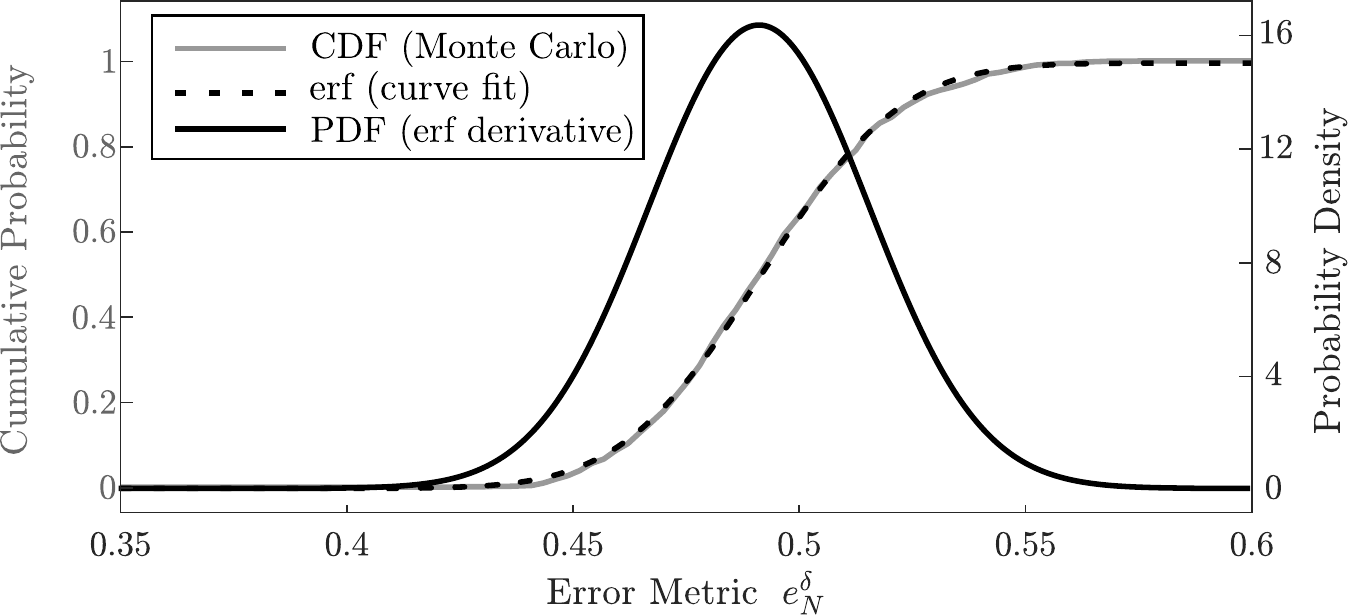}
\caption{The CDF of the error metric when robot positions are sampled from $\rho$ is approximated by Monte Carlo integration, an $\erf(\cdot)$ curve fit matches closely, and the PDF is taken as the derivative of the fitted $\erf(\cdot)$. This figure appears as Figure 4 in the authors' \cite{ICINCO2018}.}
\label{fig:error_probabilities}
\end{center}
\end{figure}

We approximate  $F_{e_N^\delta}$ using $M=1000$ Monte Carlo evaluation points; this is shown by a solid gray line in Figure \ref{fig:error_probabilities}. The numerical approximation appears to closely match a Gauss error function  as theory predicts. Therefore an analytical $\erf(\cdot)$ curve, represented by the dashed line, is fit to the data using MATLAB's least squares curve fitting routine \texttt{lsqcurvefit}. To obtain $f_{e_N^\delta}$, the analytical curve fit for $F_{e_N^\delta}$ is differentiated, and the result is also shown in Figure \ref{fig:error_probabilities}.

In addition, we calculate the mean and standard deviation of $e_N^\delta$ to be, respectively, 0.4933 and  0.02484.

%\footnote{Optionally, one could use random sampling to estimate only the mean and standard deviation of the Gaussian PDF, but we chose to estimate the entire CDF to show that it matches the predictions of theory.}
% Using the theory in the previous subsection allows us to decrease computation time by several orders of magnitude. Since we already know by Theorem \ref{thm:H} that the CDF is approximately a Gauss Error Function, finding it is a matter of fitting a function from this family to the data, which is computationally much faster than fitting a possibly arbitrary function.

\subsection{Benchmark for Stochastic Controllers}
\label{subsec:2nd benchmark}
In this subsection we describe how to use the tools developed so far to assess the performance of a given stochastic controller for a given desired distribution $\rho$, number of robots $N$, and blob radius $\delta$. We then demonstrate the method via an example. % in Subsection \ref{subsec: num ex}.

We propose using two standard statistical tests, the two-sample t- and F- tests \cite{moore2009introduction}, to assess whether the performance of the control law is comparable to sampling robot positions from the target distribution. More precisely, these tests provide criteria for the rejection of the null hypothesis that the mean and variance of the steady state values of the error metric produced by a given control law are indistinguishable from mean and variance of $f_{e_N^\delta}$. 
%These statistical test do not need the full distributions to be known, but require only the mean and variance. 
% They assume it is Gaussian
%If the t-test reveals that the means are significantly different, we suggest finding the 95\% confidence interval for the difference between the means to asses. 
This is summarized in the following procedure, which assumes that $f_{e_N^\delta}$ is approximately normal. This assumption is reasonable due to Theorem \ref{thm:H}.

\begin{itemize}
\item Step 1: Numerically find the mean and variance of $f_{e^\delta_N}$ using either of the methods described in Subsection \ref{subsec:num}.
\item Step 2: Find the mean and variance of the steady state error metric values produced by the controller.
\item Step 3: Perform two-sample t- and F-tests to compare the means and variances, respectively.
\item Step 4: Assess the results.
\begin{itemize}
\item Step 4.a: If the tests fail to refute the null hypotheses, the performance of the controller is consistent with sampling robot positions from the target distribution.
\item Step 4.b: If the two-sample t-test suggests that the means are significantly different, find the 95\% confidence interval of their difference to assess the magnitude of the controller's  performance surplus or deficiency.
\end{itemize}
\end{itemize}

Just as for the benchmark $e_{\text{rel}}$ defined in Section \ref{sec:extrema}, the first step is independent of choice of controller. Thus, if the goal is to compare several controllers, the computations of Step 1 need to be performed only once.

% The results of the statistical tests and the confidence interval demonstrate whether the error values produced by the given controller are close to those produced simply by sampling from the target distribution. Thus, we propose using these results to judge the efficacy of a given controller.

\subsubsection{Example}
\label{subsec: num ex}

We apply this method to assess the performance of the controller in \cite{li2017decentralized},  where the desired distribution is $\rho = \rho_{\text{ring}}$ (see Definition \ref{defn:ring}),  there are $N=200$ robots, and the radius is $\delta=2\text{in}$. These calculations were originally performed in the authors' \cite{ICINCO2018}.

%We approximate the CDF using $M=1000$ Monte Carlo evaluation points; this is shown by a solid gray line in Figure \ref{fig:error_probabilities}. The numerical approximation appears to closely match a Gauss Error Function ($\erf(\cdot)$, the integral of a Gaussian $G$) as theory predicts. Therefore an analytical $\erf(\cdot)$ curve, represented by the dashed line, is fit to the data using MATLAB's least squares curve fitting routine \code{lsqcurvefit}. To obtain the probability density function of the error metric, the analytical curve fit for the CDF is differentiated, and the result is also shown in Figure \ref{fig:error_probabilities}.

%With the error metric distribution now confirmed to be approximately normal, the F- and T-tests are appropriate statistical procedures for comparing the steady state error distribution to $f_{e_N^\delta}$.

Step 1, the computation of the mean and variance of $e_N^\delta$  when robot positions are sampled at random from $\rho_{\text{ring}}$, was completed in Subsection \ref{subsec:compute f F}. 
For Step 2, we use the data presented as Figure 7 of \cite{li2017decentralized} to calculate that the distribution of steady state error metric values produced by their controller  has a mean of $0.5026$ with a standard deviation of $0.02586$. These values are summarized in Table \ref{table means}.
\begin{table}[]
\caption{Summary of mean and standard deviation of error metric values for the example in Subsection \ref{subsec: num ex}.}
\label{table means}
\begin{center}
\begin{tabular}{c|c|c|}
\cline{2-3}
                                                                  & \, Mean \,    & Standard deviation \\ \hline
\multicolumn{1}{|c|}{positions sampled from $\rho_{\text{ring}}$} & 0.4933 & 0.02484            \\ \hline
\multicolumn{1}{|c|}{steady state error values}                   & 0.5026 & 0.02586            \\ \hline
\end{tabular}
\end{center}
\end{table}

Next we perform the two-sample F-test and two-sample t-test. We take the null hypothesis to be that the distribution of these error metric values is the same as $f_{e_N^\delta}$.  The results are summarized in Table \ref{table tests}.

\begin{table}[]
\caption{Summary of statistical tests for the example in Subsection \ref{subsec: num ex}.}
\label{table tests}
\begin{center}
\begin{tabular}{|l|l|}
\hline
F-statistic & 1.0831 \\ \hline
t-statistic & 8.5888 \\ \hline
\end{tabular}
\end{center}
\end{table}

The F-statistic of 1.0831 fails to refute the null hypothesis, indicating that there is no significant difference in the standard deviations. On the other hand, a two-sample t-test rejects the null hypothesis with a t-statistic of 8.5888, indicating that the steady state error is not distributed with the same population mean as $f_{e_N^\delta}$. Nonetheless, the 95\% confidence interval for the true difference between population means is computed to be merely $(0.00717,0.01141)$. This shows that the mean steady state error achieved by this controller is unlikely to exceed that of $f_{e_N^\delta}$ by more than 2.31\%. Therefore, we find the performance of the controller in \cite{li2017decentralized} to be acceptable given its stochastic nature, as the error metric values it realizes are only slightly different from those produced by sampling robot positions from the target distribution.

\begin{remark}As with $e_\text{rel}$ of Section \ref{sec:extrema}, the sentiment of this benchmark is preceded by \cite{li2017decentralized}. However, here we have presented a concrete, repeatable, and objective approach that makes this benchmark suitable for general use. In particular, we have introduced into this context the use of appropriate statistical tests for comparing two approximately normal distributions. On the other hand, \cite{li2017decentralized} relies on visual inspection, noting, ``the error values [from simulation] mostly lie between \dots the 25th and 75th percentile error values when robot configurations are randomly sampled from the target distribution''.  The fact that  $f_{e_N^\delta}$ is approximately Gaussian (according to Theorem \ref{thm:H}) not only allows for the use of these statistical tests, but also provides a way to greatly speed up computation. Indeed, the calculations in \cite{li2017decentralized}\footnote{According to the caption of Figure 2 of \cite{li2017decentralized}, the figure was generated as a histogram from 100,000 Monte Carlo samples.} took two orders of magnitude more computation than we perform in Subsection \ref{subsec:compute f F}. 
%These improvements make this error metric PDF benchmark objective and efficient, and thus suitable for common use.
\end{remark}

\section{\uppercase{Future Work}}
\label{sec:future_work}

Several open problems remain regarding the computational and theoretical aspects of our proposed performance
assessment methods. For instance, the benchmark calculations scale quickly with the size of the swarm, and therefore they
become unfeasible for a standard personal computer with swarms on the order of 1000 robots. To extend our proposed methods to larger swarms, computational techniques should be improved. Possibilities include the derivation of an analytical representation of the error metric PDF in terms of the target distribution, more sophisticated optimization algorithms to compute the extrema, and alternate formulations of the optimization problem that reduce the number of variables or introduce guarantees on the global optimality of a resulting solution. As for theoretical extensions, analysis of the shape and size of the robot blob function, as well as a rigorous understanding of the optimal
radius, are intriguing topics for further study. Specifically, if a certain control law were characterized as ``good'' with one choice of blob function, would the same conclusion be reached with a different blob function? These questions were raised by the authors' \cite{ICINCO2018}, and here we provide possible routes to answering them, as well as connections to known results in the theory (see Subsection \ref{sec:kde}).

\section{\uppercase{Conclusion}}
\label{sec:conclusion}

This work deals with the performance assessment of spatial density control of robotic swarms. We introduce a sensitive error
metric and two corresponding benchmarks, one based on the realizable extrema of the error metric and one based on its
probability density function, which provide methods for evaluating the performance of control algorithms. The parameters of
the error metric correspond to physical properties such as robot shape and size, and can therefore be
tailored to the particular system under study. The benchmark calculations then provide reference points to which the performance of different control schemes can be compared. The error metric and these benchmarks were first carefully defined in the authors' \cite{ICINCO2018}; here, we have made these definitions more transparent, and, in Subsections \ref{sec:extrema_bounds_calculation} and \ref{subsec:2nd benchmark}, we have added step-by-step methods for performing the relevant computations. This will allow swarm designers and practitioners to more easily use these benchmarks to quantitatively decide which control scheme best fits their application. As demonstrated by the examples, including the new example featured in Subsections \ref{sec:extrema_bounds_calculation} and \ref{subsec:swarm design}, these evaluation methods work well for a variety of target distributions and can be applied to judge the performance of both deterministic and stochastic control algorithms.

\section*{\uppercase{Acknowledgements}}

\noindent The authors gratefully acknowledge the support of NSF grant CMMI-1435709, NSF grant DMS-1502253, the Dept. of Mathematics at UCLA, and the Dept. of Mathematics at Harvey Mudd College. OT acknowledges support from the Charles Simonyi Endowment at the Institute for Advanced Study. The authors also thank Hao Li (UCLA) for his insightful contributions regarding the connection between this error metric and kernel density estimation.

\bibliographystyle{spmpsci}
{\small
\bibliography{example}}

\end{document}